\documentclass[aps,preprint,showpacs,amsmath,tightenlines]{revtex4}

\usepackage{bm}

\begin{document}

\title{Mirror matter admixtures in $K_{\rm L}\to\mu^+\mu^-$}

\author{G.~S\'anchez-Col\'on}

\email[]{gsanchez@mda.cinvestav.mx}

\affiliation{
Departamento de F\'{\i}sica Aplicada.\\
Centro de Investigaci\'on y de Estudios Avanzados del IPN.\\
Unidad Merida.\\
A.P. 73, Cordemex. \\
M\'erida, Yucat\'an, 97310. MEXICO.
}

\author{A.~Garc\'{\i}a}

\affiliation{
Departamento de F\'{\i}sica.\\
Centro de Investigaci\'on y de Estudios Avanzados del IPN.\\
A.P. 14-740.\\
M\'exico, D.F., 07000. MEXICO.
}

\date{\today}

\begin{abstract}

Our previous analysis on the contributions of mirror matter admixtures
in ordinary hadrons to $K_{\rm L}\to\gamma\gamma$ is extended to study
the relevance of such contributions to the $K_{\rm L}\to\mu^+\mu^-$
rare decay. The mixing angle of the admixtures previously determined
to describe the enhancement phenomenon in two body non-leptonic
decays of strange hadrons is used, along with recent results for the
description of the strong and electromagnetic interaction parts of
the transition amplitudes. We find that these admixtures give a
significant contribution with a small SU(3) breaking of only $2.8\%$,
we also find a value of $\sim -17.9^{\circ}$ for the $\eta$-$\eta'$
mixing angle consistent with some of its determinations in the
literature and a preferred negative value around $-16$ for the local
counter-term contribution $\chi_1+\chi_2$ consistent with the
existence of a unique counter-term assuming lepton universality. We
conclude that those mixings may be relevant in low energy physics and
should not be ignored.

\end{abstract}


\maketitle

Recently, we studied whether contributions of new matter may be
relevant in the rare decay $K_{\rm L}\to \gamma\gamma$~\cite{kl2g}. This study cannot
be performed in a general way and one is required to use specific
models. In particular, we applied to this decay a model we have refered
to as manifest mirror matter admixtures in ordinary hadrons~\cite{apriori}. Our
conclusion was, that indeed contributions of new physics via admixtures
of this type may be relevant in this decay and, accordingly, they must
be kept in mind when studying the description of the Standard Model (SM)
of this decay.

As is well known the two photon decay mode of $K_{\rm L}$ is closely related to
the also rare decay mode $K_{\rm L}\to\mu^+\mu^-$ and, because of this, we are
required to extend our analysis of Ref.~\cite{kl2g} to this latter mode.
This we shall do in this paper

Before proceeding, let us first review the current situation on this
decay. Its branching ratio has been gradually measured reaching very
recently quite a substantial precision, currently, ${\rm Br}(K_{\rm
L}\to\mu^+\mu^-)=(7.27\pm 0.14)\times10^{-9}$~\cite{pdg}. On the
theoretical side, the short distance contributions due in the
Standard Model to $W$ and $Z$ exchange are quite small and cannot
explain the experimental amplitude. Then, the long distance
contribution from the $2\gamma$ intermediate state is dominant. This
long distance amplitude has a large absorptive part that almost
saturates the total $K_{\rm L}\to\mu^+\mu^-$ rate (in principle, the
absorptive amplitude receives additional contributions from real
intermediate states other than two photons, such as two- and
three-pion cuts, but these are completely
negligible~\cite{martin70}). However, the dispersive part of the
$2\gamma$ contribution cannot be calculated in a model-independent
way and it is subject to various
uncertainties~\cite{isidori04,knecht99}.

Let us now proceed to apply our phenomenological model to $K_{\rm
L}\to\mu^+\mu^-$. As we mentioned above, we introduce parity and flavour
admixtures of mirror matter in ordinary mesons~\cite{apriori}, and the
$K_{\rm L}\to\mu^+\mu^-$ amplitude is assumed to be enhanced by parity and
flavour conserving amplitudes arising from the matrix elements of the ordinary
strong and electromagnetic parts of the Hamiltonian between states with
such admixtures.

The ordinary physical mesons $K^0_{\rm ph}$ and $\bar{K}^0_{\rm ph}$ with
parity and SU(3)-flavor violating admixtures are given by~\cite{apriori}
 
\[
K^0_{\rm ph} =
K^0_{\rm p} -
\frac{1}{\sqrt 2} \sigma \pi^0_{\rm p} +
\sqrt{\frac{3}{2}} \sigma \eta_{\rm 8p} +
\sqrt{\frac{2}{3}} \delta \eta_{\rm 8s} -
\frac{1}{\sqrt 3} \delta \eta_{\rm 1s} -
\frac{1}{\sqrt 2} \delta' \pi^0_{\rm s} +
\frac{1}{\sqrt 6} \delta' \eta_{\rm 8s} +
\frac{1}{\sqrt 3} \delta' \eta_{\rm 1s},
\]

\begin{equation}
\bar{K}^0_{\rm ph} =
\bar{K}^0_{\rm p} -
\frac{1}{\sqrt{2}} \sigma \pi^0_{\rm p} +
\sqrt{\frac{3}{2}} \sigma \eta_{\rm 8p} -
\sqrt{\frac{2}{3}} \delta \eta_{\rm 8s} +
\frac{1}{\sqrt{3}} \delta \eta_{\rm 1s} +
\frac{1}{\sqrt{2}} \delta' \pi^0_{\rm s} -
\frac{1}{\sqrt{6}} \delta' \eta_{\rm 8s} -
\frac{1}{\sqrt{3}} \delta' \eta_{\rm 1s}.
\label{uno}
\end{equation}

\noindent
We have used the SU(3)-phase conventions of Ref.\cite{deswart}. We recall
that the mixing angles $\sigma$, $\delta$, and $\delta'$ are the
parameters of the model, which have been determined
previously~\cite{detailed}. The subindices ${\rm s}$ and ${\rm p}$ refer
to positive and negative parity eigenstates, respectively. Notice that the
physical mesons satisfy ${\rm CP}K^0_{\rm ph}=-\bar{K}^0_{\rm ph}$ and
${\rm CP}\bar{K}^0_{\rm ph}=-K^0_{\rm ph}$.

We can form the CP-eigenstates $K_{\rm 1}$ and $K_{\rm 2}$ as

\[
K_{\rm 1_{ph}} = \frac{1}{\sqrt{2}} (K^0_{\rm ph} - \bar{K}^0_{\rm ph})
\qquad\mbox{and}\qquad
K_{\rm 2_{ph}} = \frac{1}{\sqrt{2}} (K^0_{\rm ph} + \bar{K}^0_{\rm ph}),
\]

\noindent
the $K_{\rm 1_{ph}}$ ($K_{\rm 2_{ph}}$) is an even (odd) state with respect
to CP. Here, we shall not consider CP-violation and therefore,
$|K_{\rm S,L}\rangle = |K_{1,2}\rangle$.

Substituting the expressions given in Eqs.~(\ref{uno}), we obtain,

\[
K_{\rm S_{ph}} =
K_{\rm S_p} +
\frac{1}{\sqrt{3}} (2\delta + \delta') \eta_{\rm 8s} -
\delta' \pi^0_{\rm s} -
\sqrt{\frac{2}{3}} (\delta - \delta') \eta_{\rm 1s},
\]

\begin{equation}
K_{\rm L_{ph}} =
K_{\rm L_p} -
\sigma \pi^0_{\rm p} +
\sqrt{3} \sigma \eta_{\rm 8p},
\label{tres}
\end{equation}

\noindent
where the usual definitions
$K_{\rm 1_p} = (K^0_{\rm p} - \bar{K}^0_{\rm p})/\sqrt{2}$
and
$K_{\rm 2_p} = (K^0_{\rm p} + \bar{K}^0_{\rm p})/\sqrt{2}$
were used.

As pointed out above, mirror matter admixtures in the physical mesons will
contribute to the $K_{\rm S,L}\to\mu^+\mu^-$ amplitudes via the parity and
flavour-conserving part ${\cal H}_0$ of the full Hamiltonian ${\cal H}$,
which contains the ordinary strong and electromagnetic interactions.  The
transition amplitudes will be given by the matrix elements
$\langle\mu^+\mu^-|{\cal H}_0|K_{\rm S,L_{ph}}\rangle$.  Using the above
mixings, Eqs.~(\ref{tres}), these amplitudes will have the form
$\langle\mu^+\mu^-|{\cal H}_0|K_{\rm S,L_{ph}}\rangle
=\bar{u}_{\mu^-}(p_{\mu^-}) F_{K_{\rm S,L}\mu^+\mu^-}
v_{\mu^+}(p_{\mu^+})$, where,

\begin{equation}
F_{K_{\rm S}\mu^+\mu^-} =
\frac{1}{\sqrt{3}}(2\delta+\delta')F_{\eta_{\rm 8s}\mu^+\mu^-} -
\delta' F_{\pi^0_{\rm s}\mu^+\mu^-} -
\sqrt{\frac{2}{3}}(\delta-\delta')F_{\eta_{\rm 1s}\mu^+\mu^-}
\label{cuatro}
\end{equation}

\noindent
and

\begin{equation}
F_{K_{\rm L}\mu^+\mu^-} =
-\sigma F_{\pi^0_{\rm p}\mu^+\mu^-} +
\sqrt{3}\sigma F_{\eta_{\rm 8p}\mu^+\mu^-}.
\label{cinco}
\end{equation}

\noindent
Given that $K_{\rm S}$ and $K_{\rm L}$ are ${\rm CP}=+1$ and ${\rm CP}=-1$
pure states, respectively, and because the $\mu^+\mu^-$ state is a ${\rm
C}=+1$ state, then $K_{\rm S}\to\mu^+\mu^-$ must go through a so-called
parity-violating transition while $K_{\rm L}\to\mu^+\mu^-$ goes through a
parity-conserving transition. In the first case the $\mu^+\mu^-$ final
state is ${\rm P}=+1$ while in the second one, ${\rm P}=-1$. $F_{K_{\rm
S}\mu^+\mu^-}$ and $F_{K_{\rm L}\mu^+\mu^-}$ contribute, respectively,
along the corresponding  strangeness changing and parity violating and
strangeness changing but  parity conserving amplitudes of the SM mediated
by $W^{\pm}$. However, as we  can see, from Eqs.~(\ref{cuatro}) and
(\ref{cinco}), the contributions of the mirror matter  admixtures are all
flavor and parity conserving. The additive terms on the right-hand side of
these equations involve only mirror mesons in $F_{K_{\rm S}\mu^+\mu^-}$ and
only ordinary mesons in $F_{K_{\rm L}\mu^+\mu^-}$. However, notice that
these states must carry the mass of the physical kaon $m_K$. The effective
coupling constants $F_{\pi^0_{\rm s,p} \mu^+\mu^-}$, $F_{\eta_{\rm 8s,p}
\mu^+\mu^-}$, and $F_{\eta_{\rm 1s} \mu^+\mu^-}$ correspond to the parity
and flavour conserving decay processes $\pi^0_{\rm s,p},\eta_{\rm
8s,p},\eta_{\rm 1s} \to\mu^+\mu^-$, of the $\pi^0_{\rm s,p}$, $\eta_{\rm
8s,p}$ and $\eta_{\rm 1s}$ parity and flavour eigenstates present in the
decaying physical $K_{\rm S,L_{ph}}$.

Let us now concentrate on $K_{\rm L}\to\mu^+\mu^-$,
Eq.~(\ref{cinco}). As a working hypothesis we shall assume that
the experimental branching ratio is saturated by the
contribution of such admixtures, and we shall neglect the SM
contributions. This is the same assumption we have used in our
previous work. Of course, this is an extreme assumption.
However, the reason for adopting it is that it provides a
stringent test on the above admixtures. Clearly, if a poor or
even a wrong prediction is obtained, then severe constraints on
those admixtures are imposed.

To the leading order (the fourth order) in electromagnetic
interaction and to all orders in strong interaction, the decay
amplitudes of the non-strange mesons $P=\pi^0,\eta_8$ into
$l^+l^-$ are given in terms of the purely real couplings to two
on-shell photons $F_{P\gamma\gamma}$~\cite{bergstrom82},

\begin{equation}
F_{Pl^+l^-} = 2\alpha^2m_l F_{P\gamma\gamma}R_P.
\label{F(P2mu)}
\end{equation}

\noindent
The relevant dynamics is contained in the reduced amplitudes
$R_P$.

Assuming the obvious dominance of the two photon contribution, the reduced
amplitudes $R_P=R(q^2)$ can be written as~\cite{bergstrom82}

\[
R(q^2) = \frac{2}{i\pi^2q^2}
\int d^4 k \frac{q^2k^2-(q\cdot k)^2}{[k^2+i\epsilon][(q-k)^2+i\epsilon]
[(p-k)^2-m^2_l+i\epsilon]}F(k^2,(q-k)^2)
\]

\noindent
where $q^2=m^2_P$ and $F$ is a generic and model dependent form
factor, with $F(0,0)=1$ for on-shell photons. The absorptive parts of $R_P$
are finite and model independent and are given by~\cite{drell59},

\begin{equation}
{\rm Im}R_P = \frac{\pi}{2\beta_P}
\ln\left(\frac{1-\beta_P}{1+\beta_P}\right),
\label{ImR_P}
\end{equation}

\noindent
with $\beta_P=\sqrt{1-4m^2_l/m^2_P}$. By contrast, their real parts
contain an {\it a priori} divergent $\gamma\gamma$ loop (if a constant
$F(k^2,(q-k)^2)=1$ form factor is assumed). The cure to this problem is
model dependent and proceeds either through the inclusion of non-trivial
form factors -which depend on the hadronic physics governing the
$P\to\gamma^*\gamma^*$ transition- or, in a more modern ChPT language, the
inclusion of local counter-terms to render the result
finite~\cite{savage92,ametller93}, namely,

\begin{eqnarray}
{\rm Re}R(q^2=m^2_P) &=& -\frac{\chi_1(\Lambda)+\chi_2(\Lambda)}{4}
-\frac{5}{2} +3\ln\left(\frac{m_l}{\Lambda}\right)
\nonumber\\
&& +\frac{1}{4\beta_P}\ln^2\left(\frac{1-\beta_P}{1+\beta_P}\right)
+\frac{\pi^2}{12\beta_P}
+\frac{1}{\beta_P}{\rm Li}_2\left(\frac{\beta_P-1}{\beta_P+1}\right).
\label{ReR_P}
\end{eqnarray}

\noindent
The explicit logarithmic dependence on $\Lambda$ reflects the ultraviolet
divergence of the loop and cancels with the inclusion of the local
counter-terms contribution $\chi_1(\Lambda)+\chi_2(\Lambda)$.

From Eqs.~(\ref{cinco})-(\ref{ReR_P}), the magnitude of the $K_{\rm
L}\to\mu^+\mu^-$ amplitude in the mirror matter admixtures context, is
given in terms of the $F_{\pi^0_{\rm p}\gamma\gamma}$ and $F_{\eta_{\rm
8p}\gamma\gamma}$ decay amplitudes, the mixing angle $\sigma$, and the
local contribution $\chi_1(\Lambda)+\chi_2(\Lambda)$. It is explicitly
given by

\begin{eqnarray}
|F_{K_{\rm L}\mu^+\mu^-}| &=&
2\alpha^2 m_{\mu} \, |-\sigma F_{\pi^0_{\rm p}\gamma\gamma} +
\sqrt{3}\sigma F_{\eta_{\rm 8p}\gamma\gamma}|
\nonumber\\
&\times& \left\{ \left[ -\frac{\chi_1(\Lambda)+\chi_2(\Lambda)}{4}-\frac{5}{2}
+3\ln\left(\frac{m_{\mu}}{\Lambda}\right)
+\frac{1}{4\beta_K}\ln^2\left(\frac{1-\beta_K}{1+\beta_K}\right)
\right.\right.
\nonumber\\
&& \left.\left. +\frac{\pi^2}{12\beta_K} +\frac{1}{\beta_K}{\rm
Li}_2\left(\frac{\beta_K-1}{\beta_K+1}\right)\right]^2+\left[\frac{\pi}{2\beta_K}
\ln\left(\frac{1-\beta_K}{1+\beta_K}\right)\right]^2
\right\}^{1/2},
\label{FK2mu} \end{eqnarray}

\noindent
where use has been made of $\beta_{\pi^0_{\rm
p}}= \beta_{\eta_{\rm 8p}}= \beta_K= \sqrt{1-4m^2_{\mu}/m^2_K}$,
because $\pi^0_{\rm p}$ and $\eta_{\rm 8p}$ share
the mass $m_K$, as mentioned before.

To be able of perform a numerical application, we shall
include in our analysis the experimentally observed processes
$K_L,\eta,\eta' \to\gamma\gamma$. $K_L\to \gamma\gamma$ in the mirror
matter admixtures context is also related to $\pi^0\to \gamma\gamma$
and $\eta_8\to \gamma\gamma$ by a relation analogous to
Eq.~(\ref{cinco})~\cite{kl2g},

\begin{equation}
F_{K_{\rm L}\gamma\gamma} =
-\sigma F_{\pi^0_{\rm p}\gamma\gamma} +
\sqrt{3}\sigma F_{\eta_{\rm 8p}\gamma\gamma}.
\label{kl2g}
\end{equation}

\noindent
Notice that in this context, $F_{K_{\rm L}\gamma\gamma}$ and in
consequence $F_{K_{\rm L}\mu^+\mu^-}$ vanish in the strong flavour
SU(3) symmetry limit (U-spin invariance)~\cite{kl2g}.Thus, if we define

\begin{equation}
\Delta =\frac{F_{\eta_8\gamma\gamma}}{F_{\pi^0\gamma\gamma}/\sqrt{3}},
\label{su32g}
\end{equation}

\noindent
then in the symmetry limit one has $\Delta=1$.

Concerning the $\eta,\eta' \to\gamma\gamma$ processes, it has been
established that in general two angles are necessary to describe the
$\eta$-$\eta'$ mixing scheme. One cannot assume that the same rotation
applies to the octet-singlet states and to their decay constants. For a
review see Ref.~\cite{feldmann00}. However, we shall use this mixing
scheme only at the amplitude level and in this case only one mixing angle
appears~\cite{cao99}. In this respect, it should be clear that we are not
making the questionable assumption that only one mixing angle is used both
for the states and the decay constants. Then, following Ref.~\cite{cao99},
we introduce the rotation

\[
\left(
\begin{array}{c}
\eta_{\rm p} \\ \eta'_{\rm p}
\end{array}
\right)
=
\left(
\begin{array}{cc}
\cos{\theta_{\rm p}} & -\sin{\theta_{\rm p}} \\
\sin{\theta_{\rm p}} & \cos{\theta_{\rm p}}
\end{array}
\right)
\left(
\begin{array}{c}
\eta_{\rm 8p}\\
\eta_{\rm 1p}
\end{array}
\right)
\]

\noindent
and this leads at the amplitude level to

\begin{equation}
F_{\eta_{\rm p}\gamma\gamma}
=
\cos{\theta_{\rm p}}F_{\eta_{\rm 8p}\gamma\gamma}
-\sin{\theta_{\rm p}}F_{\eta_{\rm 1p}\gamma\gamma},
\label{ochoa}
\end{equation}

\begin{equation}
F_{\eta'_{\rm p}\gamma\gamma}
=
\sin{\theta_{\rm p}}F_{\eta_{\rm 8p}\gamma\gamma}
+\cos{\theta_{\rm p}} F_{\eta_{\rm 1p}\gamma\gamma}.
\label{ochob}
\end{equation}

The experimental data we shall use come from Ref.~\cite{pdg}. The
corresponding experimental values for $|F_{K_{\rm L}\mu^+\mu^-}|$ and
$F_{P\gamma\gamma}$ ($P=K_{\rm L},\pi^0,\eta,\eta'$) are displayed in
Table~\ref{table1}. They are obtained using

\begin{equation}
|F_{P l^+ l^-}| =
\left[\frac{8\pi}{m_P\beta_P}\Gamma(P\to l^+ l^-)\right]^{1/2}
\label{9a}
\end{equation}

\noindent
and

\begin{equation}
F_{P\gamma\gamma} =
\pm\frac{2}{\alpha}\left[\frac{1}{\pi m^3_P}
\Gamma(P\to\gamma\gamma)\right]^{1/2}.
\label{P2g}
\end{equation}

Also, the SU(3) symmetry relation, Eq.~(\ref{su32g}), should be obeyed
within a reasonable breaking of SU(3). We shall assume in what follows
that Eq.~(\ref{su32g}) is obeyed within an uncertainty of about $10\%$.

Of the values for the mixing angles of the mirror matter admixtures
obtained previously~\cite{detailed}, we shall only need
$\sigma=(4.9\pm2.0)\times10^{-6}$. We do not quote the values of the other
two mixing angles because we shall not use them here.

Experimental values for the local contribution
$\chi_1(\Lambda)+\chi_2(\Lambda)$ can be determined from Eq.~(\ref{ReR_P})
by using Eq.~(\ref{ImR_P}) and the \lq\lq reduced" ratio

\begin{equation}
\frac{{\rm Br}(P\to l^+ l^-)}{{\rm Br}(P\to\gamma\gamma)}=
2\beta_P\left(\frac{\alpha m_l}{\pi m_P}\right)^2 ({\rm Re}^2R_P + {\rm
Im}^2R_P).
\label{B}
\end{equation}

\noindent
The present experimental $\eta\to\mu^+\mu^-$
branching ratio~\cite{pdg} requires a counter-term (for
$\Lambda=m_{\rho}=775.8\pm 0.5\ {\rm MeV}$) given by

\begin{subequations}
\label{chis}
\begin{equation}
\chi_1(m_{\rho})+\chi_2(m_{\rho})=
-6.8\pm 3.5,
\label{chisa}
\end{equation}
\begin{equation}
\chi_1(m_{\rho})+\chi_2(m_{\rho})=
-31.9\pm 3.5.
\label{chisb}
\end{equation}
\end{subequations}

\noindent
In turn, the less precise $\pi^0\to e^+e^-$
available experimental data~\cite{pdg} translate
correspondingly into

\begin{subequations}
\label{chise}
\begin{equation}
\chi_1(m_{\rho})+\chi_2(m_{\rho})=
69.5\pm 6.6,
\label{chisea}
\end{equation}
\begin{equation}
\chi_1(m_{\rho})+\chi_2(m_{\rho})=
-10.0\pm 6.6.
\label{chiseb}
\end{equation}
\end{subequations}

\noindent
Values~(\ref{chisa}) and (\ref{chisea}) correspond to the negative root of
${\rm Re}R_P$ in Eq.~(\ref{B}) and values~(\ref{chisb}) and (\ref{chiseb})
to the positive one. The values~(\ref{chisa}) and (\ref{chiseb}) are
favored because they are consistent with the existence of a unique
counter-term assuming lepton universality and they are also in coincidence
with the predictions of the resonance saturation
hypothesis~\cite{ametller93} and Lowest Meson Dominance in large-$N_c$ QCD
models~\cite{knecht99}. On these grounds, the values~(\ref{chisb}) and
(\ref{chisea}) should be discarded. In our analysis we shall consider each
value of Eqs.~(\ref{chis}) and (\ref{chise}) separately.

From the above formulation of the problem, we are now in a position to make
a prediction for the $K_{\rm L}\to\mu^+\mu^-$. As we mentioned above, we
shall use the $K_L,\pi^0,\eta,\eta' \to\gamma\gamma$ decay amplitudes as
constraints. The way to proceed is to form a $\chi^2$ function and fit it.
This $\chi^2$ contains eight summands, corresponding to the five
experimental amplitudes, the allowed range of SU(3) breaking, the range
for $\sigma$ determined from our analysis of strange hadron two-body
non-leptonic decays, and the local contribution to the $P\to l^+l^-$ decay
amplitudes of Eqs.~(\ref{chis}) and (\ref{chise}). Then the six quantities,
$F_{\pi^0_{\rm p}\gamma\gamma}$, $F_{\eta_{\rm 8p}\gamma\gamma}$,
$F_{\eta_{\rm 1p}\gamma\gamma}$, $\sigma$, $\chi_1(m_{\rho})+
\chi_2(m_{\rho})$, and $\theta_{\rm p}$, are allowed to minimize this
$\chi^2$. All the two possible signs in front of each $F_{P\gamma\gamma}$
($P=K_{\rm L},\pi^0,\eta,\eta'$) amplitude must be explored. Our best
results are displayed in Tables~\ref{table1} and \ref{table2}.

Looking through Tables~\ref{table1} and~\ref{table2}, one can observe
several features. In all the four cases considered the experimental
$K_{\rm L}\to\mu^+\mu^-$ amplitude is well reproduced, the other four
experimental amplitudes are also well reproduced, the ranges obtained for
$\sigma$ overlap neatly with the one of Ref.~\cite{detailed}, the
$\eta$-$\eta'$ mixing angle $\theta_{\rm p}$ is stable and consistent with
values obtained recently~\cite{cao99,yeh02feldmann98}, and finally the
SU(3) symmetry breaking $\Delta$ parameter is also stable and
corresponds to quite a small breaking ($\sim 2.8\%$). The important
differences that can be observed in these tables come from the
$\chi_1(m_{\rho})+ \chi_2(m_{\rho})$ local contribution. Its positive
value of~(\ref{chisea})  is clearly excluded. A negative value around
$-16$ is preferred; however, the range down to $-25$ may still be
acceptable. This last is consistent  with a unique counter-term in ChPT.

The above analysis shows that, in the framework of manifest mirror
matter admixtures, new physics may give relevant contributions to the
description of $K_{\rm L}\to\mu^+\mu^-$. At his point we should notice an
important parallelism with $K_{\rm L}\to\gamma\gamma$. The angle $\sigma$ previously
determined from strange hadron two-body non-leptonic decays predicts
an estimate for the $K_{\rm L}\to\mu^+\mu^-$ branching ratio larger
than its experimental value, the SU(3) symmetry limit cancellation of
the transition amplitude corresponding to Eq.~(\ref{su32g}) then
requires the small symmetry breaking of $2.8\%$ to reproduce the
experimental value. This is the same mechanism that in the $K_{\rm
L}\to\gamma\gamma$ case led to small symmetry breaking, too.

Let us conclude with an important remark. The real constraint one can
obtain from studies such as the above one and our previous ones is on
the existence of new forms of matter, specifically of mirror matter. Its
contributions in low enery physics are relevant only is as much as the
SM contributions leave room for them to be observed. At present the
uncertainties in the determination of the SM contributions in this
realm of physics do allow for new physics to be observed there. However,
if in the future it were to be the case that the SM leaves no room for
other effects, then one should conclude that manifest mirror matter, if
it exists, can only be found very far away. According to the lower bound
established in Ref.~\cite{bound}, it would be found above $10^6$~GeV.

We would like to thank CONACyT (M\'exico) for partial support.

\begin{table}

\caption{ Experimentally observed, predicted values, and $\Delta\chi^2$
contributions of the $K_{\rm L}\to\mu^+\mu^-$ and $K_{\rm
L},\pi^0,\eta,\eta'\to\gamma\gamma$ decay amplitudes for each one of the
local counter terms values of Eq.~(\ref{chis}). Only the magnitudes of the
experimental values are displayed, the signs for the predictions of these
amplitudes correspond to the ones obtained in our best fit. In each case,
the values obtained for the parameters of the fit along with the
constraints imposed (where applicable) and their $\Delta\chi^2$
contribution are also displayed. All the $2\gamma$ decay amplitudes are in
${\rm MeV}^{-1}$. $\Delta$ gives the magnitude of SU(3) breaking and its
corresponding $\Delta\chi^2$ is with respect to the assumed $10\%$ breaking.
The total $\chi^2$ are displayed in parenthesis in the left column.
\label{table1} }

\begin{ruledtabular}

\begin{tabular}{c|c|c|c|c}

$\chi_1(m_{\rho})+\chi_2(m_{\rho})$ & Decay & Experiment &
Prediction & $\Delta\chi^2$ \\

\hline

$-6.8\pm 3.5$ & $K_{\rm L}\to\mu^+\mu^-$ & $(2.270\pm
0.024)\times 10^{-12}$ & $2.287\times10^{-12}$ & $0.50$ \\
 
& $K_{\rm L}\to\gamma\gamma$ & $(3.814\pm 0.027)\times10^{-11}$ &
$3.801\times 10^{-11}$ & $0.23$ \\

& $\pi^0\to\gamma\gamma$ & $(2.744\pm 0.098)\times 10^{-4}$ &
$-2.744\times10^{-4}$ & $\sim 10^{-6}$ \\

& $\eta\to\gamma\gamma$ & $(2.720\pm 0.074)\times 10^{-4}$ &
$-2.720\times10^{-4}$ & $\sim 10^{-14}$ \\

& $\eta'\to\gamma\gamma$ & $(3.41\pm 0.18)\times10^{-4}$ &
$-3.41\times 10^{-4}$ & $\sim 10^{-14}$ \\

\cline{2-5}

($\chi^2= 7.32$) & Parameter & Constraint & Prediction & $\Delta\chi^2$ \\

\cline{2-5}

& $F_{\pi^0_{\rm p}\gamma\gamma}$ & $(2.744\pm 0.098)\times 10^{-4}$ &
$(-2.744\pm 0.098)\times 10^{-4}$ & $\sim 10^{-6}$ \\

& $F_{\eta_{\rm 8p}\gamma\gamma}$ & --- & $(-1.540\pm 0.060)\times
10^{-4}$ & --- \\

& $F_{\eta_{\rm 1p}\gamma\gamma}$ & --- & $(-4.08\pm 0.16)\times
10^{-4}$ & --- \\

& $\sigma$ & $(4.9\pm 2.0)\times 10^{-6}$ & $(5.0\pm 2.0)\times
10^{-6}$ & $\sim 10^{-3}$ \\

& $\chi_1(m_{\rho})+\chi_2(m_{\rho})$ & $-6.8\pm 3.5$ & $-15.7\pm 1.1$
& $6.47$ \\

& $\theta_{\rm p}$ & --- & $(-17.9\pm 1.5)^\circ$ & --- \\

& $\Delta$ & $1.00 \pm 0.10$ &
$0.97 $ & $0.09$ \\

\hline

$\chi_1(m_{\rho})+\chi_2(m_{\rho})$ & Decay & Experiment &
Prediction & $\Delta\chi^2$ \\

\hline

$-31.9\pm 3.5$ & $K_{\rm L}\to\mu^+\mu^-$ & $(2.270\pm 0.024)\times
10^{-12}$ & $2.284\times 10^{-12}$ & $0.34$ \\
 
& $K_{\rm L}\to\gamma\gamma$ & $(3.814\pm 0.027)\times 10^{-11}$ &
$3.804\times 10^{-11}$ & $0.14$ \\

& $\pi^0\to\gamma\gamma$ & $(2.744\pm 0.098)\times 10^{-4}$ &
$-2.744\times 10^{-4}$ & $\sim 10^{-6}$ \\

& $\eta\to\gamma\gamma$ & $(2.720\pm 0.074)\times 10^{-4}$ &
$-2.720\times 10^{-4}$ & $\sim 10^{-14}$ \\

& $\eta'\to\gamma\gamma$ & $(3.41\pm 0.18)\times 10^{-4}$ &
$-3.41\times 10^{-4}$ & $\sim 10^{-14}$ \\

\cline{2-5}

($\chi^2= 4.23$) & Parameter & Constraint & Prediction & $\Delta\chi^2$ \\

\cline{2-5}

& $F_{\pi^0_{\rm p}\gamma\gamma}$ & $(2.744\pm 0.098)\times 10^{-4}$ &
$(-2.744\pm 0.098)\times10^{-4}$ & $\sim 10^{-6}$ \\

& $F_{\eta_{\rm 8p}\gamma\gamma}$ & --- & $(-1.540\pm
0.060)\times10^{-4}$ & --- \\

& $F_{\eta_{\rm 1p}\gamma\gamma}$ & --- & $(-4.08\pm
0.16)\times10^{-4}$ & --- \\

& $\sigma$ & $(4.9\pm 2.0)\times 10^{-6}$ & $(5.0\pm
2.0)\times10^{-6}$ & $\sim 10^{-3}$ \\

& $\chi_1(m_{\rho})+\chi_2(m_{\rho})$ & $-31.9\pm 3.5$ & $-25.2\pm
1.2$ & $3.68$ \\

& $\theta_{\rm p}$ & --- & $(-17.9\pm 1.5)^\circ$ & --- \\

& $\Delta$ & $1.00 \pm 0.10$ &
$0.97 $ & $0.09$ \\

\end{tabular}

\end{ruledtabular}

\end{table}

\begin{table}

\caption{ Experimentally observed, predicted values, and $\Delta\chi^2$
contributions of the $K_{\rm L}\to\mu^+\mu^-$ and $K_{\rm
L},\pi^0,\eta,\eta'\to\gamma\gamma$ decay amplitudes for each one of the
local counter terms values of Eq.~(\ref{chise}). Only the magnitudes of
the experimental values are displayed, the signs for the predictions of
these amplitudes correspond to the ones obtained in our best fit. In each
case, the values obtained for the parameters of the fit along with the
constraints imposed (where applicable) and their $\Delta\chi^2$
contribution are also displayed. All the $2\gamma$ decay amplitudes are in
${\rm MeV}^{-1}$. $\Delta$ gives the magnitude of SU(3) breaking and its
corresponding $\Delta\chi^2$ is with respect to the assumed $10\%$ breaking.
The total $\chi^2$ are displayed in parenthesis in the left column.
\label{table2}
}

\begin{ruledtabular}

\begin{tabular}{c|c|c|c|c}

$\chi_1(m_{\rho})+\chi_2(m_{\rho})$ & Decay & Experiment &
Prediction & $\Delta\chi^2$ \\

\hline

$69.5\pm 6.6$ & $K_{\rm L}\to\mu^+\mu^-$ & $(2.270\pm
0.024)\times10^{-12}$ & $2.308\times10^{-12}$ & $2.51$ \\
 
& $K_{\rm L}\to\gamma\gamma$ & $(3.814\pm 0.027)\times10^{-11}$ &
$3.785\times10^{-11}$ & $1.15$ \\

& $\pi^0\to\gamma\gamma$ & $(2.744\pm 0.098)\times10^{-4}$ &
$-2.744\times10^{-4}$ & $\sim 10^{-5}$ \\

& $\eta\to\gamma\gamma$ & $(2.720\pm 0.074)\times10^{-4}$ &
$-2.720\times10^{-4}$ & $\sim 10^{-12}$ \\

& $\eta'\to\gamma\gamma$ & $(3.41\pm 0.18)\times10^{-4}$ &
$-3.41\times10^{-4}$ & $\sim 10^{-13}$ \\

\cline{2-5}

($\chi^2= 166.02$) & Parameter & Constraint & Prediction $(\pm 1\sigma)$ & $\Delta\chi^2$ \\

\cline{2-5}

& $F_{\pi^0_{\rm p}\gamma\gamma}$ & $(2.744\pm 0.098)\times10^{-4}$ &
$(-2.744\pm 0.098)\times10^{-4}$ & $\sim 10^{-5}$ \\

& $F_{\eta_{\rm 8p}\gamma\gamma}$ & --- & $(-1.540\pm
0.060)\times10^{-4}$ & --- \\

& $F_{\eta_{\rm 1p}\gamma\gamma}$ & --- & $(-4.08\pm
0.16)\times10^{-4}$ & --- \\

& $\sigma$ & $(4.9\pm 2.0)\times10^{-6}$ & $(5.0\pm
2.0)\times10^{-6}$ & $\sim 10^{-3}$ \\

& $\chi_1(m_{\rho})+\chi_2(m_{\rho})$ & $69.5\pm 6.6$ & $-14.6\pm 0.9$
& $162.37$ \\

& $\theta_{\rm p}$ & --- & $(-17.9\pm 1.5)^\circ$ & --- \\

& $\Delta$ & $1.00 \pm 0.10$ &
$0.97 $ & $0.09$ \\

\hline

$\chi_1(m_{\rho})+\chi_2(m_{\rho})$ & Decay & Experiment &
Prediction & $\Delta\chi^2$ \\

\hline

$-10.0\pm 6.6$ & $K_{\rm L}\to\mu^+\mu^-$ & $(2.270\pm
0.024)\times10^{-12}$ & $2.274\times 10^{-12}$ & $0.03$ \\
 
& $K_{\rm L}\to\gamma\gamma$ & $(3.814\pm 0.027)\times 10^{-11}$ &
$3.811\times 10^{-11}$ & $0.01$ \\

& $\pi^0\to\gamma\gamma$ & $(2.744\pm 0.098)\times 10^{-4}$ &
$-2.744\times 10^{-4}$ & $\sim 10^{-5}$ \\

& $\eta\to\gamma\gamma$ & $(2.720\pm 0.074)\times10^{-4}$ &
$-2.720\times 10^{-4}$ & $\sim 10^{-12}$ \\

& $\eta'\to\gamma\gamma$ & $(3.41\pm 0.18)\times10^{-4}$ &
$-3.41\times 10^{-4}$ & $\sim 10^{-13}$ \\

\cline{2-5}

($\chi^2= 1.13$) & Parameter & Constraint & Prediction $(\pm 1\sigma)$ &
$\Delta\chi^2$ \\

\cline{2-5}

& $F_{\pi^0_{\rm p}\gamma\gamma}$ & $(2.744\pm 0.098)\times 10^{-4}$ &
$(-2.744\pm 0.098)\times 10^{-4}$ & $\sim 10^{-5}$ \\

& $F_{\eta_{\rm 8p}\gamma\gamma}$ & --- & $(-1.540\pm
0.060)\times 10^{-4}$ & --- \\

& $F_{\eta_{\rm 1p}\gamma\gamma}$ & --- & $(-4.08\pm
0.16)\times 10^{-4}$ & --- \\

& $\sigma$ & $(4.9\pm 2.0)\times 10^{-6}$ & $(5.0\pm
2.0)\times 10^{-6}$ & $\sim 10^{-3}$ \\

& $\chi_1(m_{\rho})+\chi_2(m_{\rho})$ & $-10.0\pm 6.6$ & $-16.6\pm
1.6$ & $1.00$ \\

& $\theta_{\rm p}$ & --- & $(-17.9\pm 1.5)^\circ$ & --- \\

& $\Delta$ & $1.00 \pm 0.10$ &
$0.97 $ & $0.09$ \\

\end{tabular}

\end{ruledtabular}

\end{table}

\end{document}